\documentclass[12pt,letterpaper]{article}

\usepackage{amsfonts}
\usepackage{amssymb}
\usepackage{amsmath}
\usepackage{dsfont}
\usepackage{hyperref}
\usepackage{slashed}
\usepackage{empheq}
\usepackage{multirow}
\usepackage{fancyhdr}
\usepackage{appendix}
\usepackage{subfigure}

\usepackage{amsmath,amssymb,amsfonts,amsthm,dcolumn}
\usepackage{amscd,amssymb,epsf}
\usepackage{bbm}
\usepackage{fancyhdr}

\def \ov {\over}

\newcounter{subequation}[equation]

\newcommand{\be}{\begin{equation}}
\newcommand{\ee}{\end{equation}}
\newcommand{\eel}[1]{\label{#1}\end{equation}}
\newcommand{\bea}{\begin{eqnarray}}
\newcommand{\eea}{\end{eqnarray}}
\newcommand{\eeal}[1]{\label{#1}\end{eqnarray}}

\makeatletter

\def\thesubequation{\theequation\@alph\c@subequation}
\def\@subeqnnum{{\rm (\thesubequation)}}
\def\slabel#1{\@bsphack\if@filesw {\let\thepage\relax
   \xdef\@gtempa{\write\@auxout{\string
      \newlabel{#1}{{\thesubequation}{\thepage}}}}}\@gtempa
   \if@nobreak \ifvmode\nobreak\fi\fi\fi\@esphack}
\def\subeqnarray{\stepcounter{equation}
\let\@currentlabel=\theequation\global\c@subequation\@ne
\global\@eqnswtrue \global\@eqcnt\z@\tabskip\@centering\let\\=\@subeqncr

$$\halign to \displaywidth\bgroup\@eqnsel\hskip\@centering
  $\displaystyle\tabskip\z@{##}$&\global\@eqcnt\@ne
  \hskip 2\arraycolsep \hfil${##}$\hfil
  &\global\@eqcnt\tw@ \hskip 2\arraycolsep
  $\displaystyle\tabskip\z@{##}$\hfil
   \tabskip\@centering&\llap{##}\tabskip\z@\cr}
\def\endsubeqnarray{\@@subeqncr\egroup
                     $$\global\@ignoretrue}
\def\@subeqncr{{\ifnum0=`}\fi\@ifstar{\global\@eqpen\@M
    \@ysubeqncr}{\global\@eqpen\interdisplaylinepenalty \@ysubeqncr}}
\def\@ysubeqncr{\@ifnextchar [{\@xsubeqncr}{\@xsubeqncr[\z@]}}
\def\@xsubeqncr[#1]{\ifnum0=`{\fi}\@@subeqncr
   \noalign{\penalty\@eqpen\vskip\jot\vskip #1\relax}}
\def\@@subeqncr{\let\@tempa\relax
    \ifcase\@eqcnt \def\@tempa{& & &}\or \def\@tempa{& &}
      \else \def\@tempa{&}\fi
     \@tempa \if@eqnsw\@subeqnnum\refstepcounter{subequation}\fi
     \global\@eqnswtrue\global\@eqcnt\z@\cr}
\let\@ssubeqncr=\@subeqncr
\@namedef{subeqnarray*}{\def\@subeqncr{\nonumber\@ssubeqncr}\subeqnarray}

\@namedef{endsubeqnarray*}{\global\advance\c@equation\m@ne
                           \nonumber\endsubeqnarray}

\makeatletter \@addtoreset{equation}{section} \makeatother
\renewcommand{\theequation}{\thesection.\arabic{equation}}

\catcode`\@=11

\newcount\hour
\newcount\minute
\newtoks\amorpm \hour=\time\divide\hour by 60\minute
=\time{\multiply\hour by 60 \global\advance\minute by-\hour}
\edef\standardtime{{\ifnum\hour<12 \global\amorpm={am}
        \else\global\amorpm={pm}\advance\hour by-12 \fi
        \ifnum\hour=0 \hour=12 \fi
        \number\hour:\ifnum\minute<10
        0\fi\number\minute\the\amorpm}}
\edef\militarytime{\number\hour:\ifnum\minute<10 0\fi\number\minute}

\def\draftlabel#1{{\@bsphack\if@filesw {\let\thepage\relax
   \xdef\@gtempa{\write\@auxout{\string
      \newlabel{#1}{{\@currentlabel}{\thepage}}}}}\@gtempa
   \if@nobreak \ifvmode\nobreak\fi\fi\fi\@esphack}
        \gdef\@eqnlabel{#1}}
\def\@eqnlabel{}
\def\@vacuum{}
\def\marginnote#1{}
\def\draftmarginnote#1{\marginpar{\raggedright\scriptsize\tt#1}}
\def\draft{
        \pagestyle{plain}
        \overfullrule=2pt
        \oddsidemargin -.5truein
        \def\@oddhead{\sl \phantom{\today\quad\militarytime} \hfil
        \smash{\Large\sl DRAFT} \hfil \today\quad\militarytime}
        \let\@evenhead\@oddhead
        \let\label=\draftlabel
        \let\marginnote=\draftmarginnote
        \def\ps@empty{\let\@mkboth\@gobbletwo
        \def\@oddfoot{\hfil \smash{\Large\sl DRAFT} \hfil}
        \let\@evenfoot\@oddhead}

\def\@eqnnum{(\theequation)\rlap{\kern\marginparsep\tt\@eqnlabel}
        \global\let\@eqnlabel\@vacuum}  }

\renewcommand{\theequation}{\thesection.\arabic{equation}}
\renewcommand{\thefootnote}{\fnsymbol{footnote}}

\def\appendix#1{
  \addtocounter{section}{-3}
  \setcounter{equation}{0}
  \renewcommand{\thesection}{\Alph{section}}
  \section*{Appendix \thesection\protect\indent \parbox[t]{11.15cm}
  {#1} }
  \addcontentsline{toc}{section}{Appendix \thesection\ \ \ #1}
  }
\def \ov {\over}

\def\O{\Omega}

\def\be{\begin{equation}}
\def\ee{\end{equation}}

\newcommand{\bal}{\begin{align}}
\newcommand{\ean}{\end{align}}

\begin{document}

\date{}

\begin{titlepage}

\hfill MCTP-11-19\\

\begin{center}

{\Large \bf Analytic Non-integrability in String Theory}

\vskip .7 cm

\vskip 1 cm

{\large   Pallab Basu${}^{1,2}$ and Leopoldo A. Pando Zayas${}^3$}

\end{center}

\vskip .4cm
\centerline{ \it ${}^1$ Center For Quantum Spacetime, Sogang University}
\centerline{\it Shinsu-dong 1, Mapo-gu, Seoul, South Korea}
\centerline{\it ${}^2$ Department of Physics and Astronomy, University of Kentucky}
\centerline{\it  Lexington, KY 40506, USA}

\vskip .4cm \centerline{\it ${}^3$ Michigan Center for Theoretical
Physics}
\centerline{ \it Randall Laboratory of Physics, The University of
Michigan}
\centerline{\it Ann Arbor, MI 48109-1120}

\vskip .4cm
\centerline{ \it }
\centerline{\it  }

\vskip 1 cm

\vskip 1.5 cm

\begin{abstract}
Using analytic techniques developed for Hamiltonian dynamical systems we show that a certain classical string configurations in $AdS_5\times X_5$ with $X_5$ in a large class of Einstein spaces, is non-integrable. This answers the question of integrability of string on such backgrounds in the negative. We consider a string localized in the center of $AdS_5$ that winds around two circles in the manifold $X_5$.
\end{abstract}

\end{titlepage}
\setcounter{page}{1} \renewcommand{\thefootnote}{\arabic{footnote}}
\setcounter{footnote}{0}

\def \N{{\cal N}}
\def \ov {\over}

\section{Introduction}

Chaotic motion has been one of the most studied aspects of nonlinear dynamical systems as its application extend to many areas \cite{Ott,Hilborn,jcsprott}. Although its mathematical roots date back to Poincar\`e and the three-body problem, it was really during the last part of the XX century when its study flourished largely thanks to new advances in computing power. Naturally, under the shadow of quantum mechanics it is logical to try to understand the quantum properties in systems whose classical limit is chaotic, this area has become known as quantum chaos \cite{gutzwiller-quantum-chaos}. In the context of the AdS/CFT correspondence \cite{Maldacena:1997re,Witten:1998qj,Gubser:1998bc, Aharony:1999ti}, there is a particularly special chance to understand some of these questions as we have a setting in which the classical regime of a theory is dual to the highly quantum regime of another. Understanding classical chaos and the corresponding quantization in the context of string theory provides a new framework with enhanced interpretational opportunities.

The simplest version of the AdS/CFT correspondence \cite{Maldacena:1997re,Witten:1998qj,Gubser:1998bc, Aharony:1999ti} states a complete equivalence between strings on $AdS_5\times S^5$ with Ramond-Ramond fluxes and ${\cal N}=4$ supersymmetric Yang-Mills (SYM) with gauge group $SU(N)$. Chaotic behavior of some classical configurations of strings in the  context of the AdS/CFT has been recently established for several interesting string theory backgrounds: ring strings in the Schwarzschild black hole in asymptotically $AdS_5$ backgrounds \cite{Zayas:2010fs}, strings in the AdS soliton background \cite{Basu:2011dg} and in $AdS_5\times T^{1,1}$ \cite{Basu:2011di}.

To determine  if a system is integrable the standard route is to find the integrals of motion and to show that there are as many as the number of degrees of freedom. Alternately, one can use Kolmogorov-Arnold-Moser  (KAM) theorem to show that the system is non-integrable. In general the question of integrability is settled through a numerical analysis of the system \cite{Ott,Hilborn,jcsprott}. Over the last decades an analytical approach has been developed to determine whether a system in integrable. In particular, some powerful results due to Ziglin \cite{ZiglinI, ZiglinII} and further refined by Morales-Ruiz and Ramis \cite{Morales-Ruiz} turn the question of integrability of some simple systems into an algorithmic process. In this paper we study a large class of systems that appear in string theory. In a sense we generalize some of our previous results for classical strings on $AdS_5\times T^{1,1}$ \cite{Basu:2011di} to include more general backgrounds of the form $AdS_5\times X_5$, where $X_5$ is in a general class of five-dimensional Einstein spaces admitting a $U(1)$ fibration.

The rest of the paper is organized as follows. In section \ref{sec:Analytic} we review the main techniques in the theory of dynamical system that allow us to prove analytically that a large class of string configurations in string theory are non-integrable. Namely, in section \ref{sec:Analytic} we state the main idea, definitions and results that form the core of the approach. In section \ref{sec:Models} we introduce the models that we are discussing and apply the machinery reviewed in section  \ref{sec:Analytic}; we establish that a specific string configuration is not integrable in the Liouville sense for a large class of string backgrounds. We also show in  \ref{sec:Models}, as an example, that the wrapped string in  $AdS_5\times S^5$ is integrable using the same algorithm.  We conclude in section \ref{sec:conclusions} with some general comments about the ingredients required for our analysis and point to some interesting future problems.

\section{Analytic Non-integrability}\label{sec:Analytic}
Let us briefly review the main statements of the area of analytic\footnote{By analytic we mean meromorphic. A meromorphic function on an open subset D of the complex plane is a function that is holomorphic on all D except a set of isolated points, which are poles for the function.} non-integrability  \cite{Fomenko,Morales-Ruiz,Goriely}. The central place in the study of integrability and non-integrability of dynamical systems is occupied by ideas developed in the context of the KAM theory. The KAM theorem describes how an integrable system reacts to small deformations. The lost of integrability is readily characterized by the resonant properties of the corresponding phase space tori, describing integrals of motion in the action-angle variables. These ideas were already present in Kovalevskaya's work but were made precise in the context of KAM theory.

The general basis for proving nonintegrability of a system of differential equations $\dot{\vec{x}}=\vec{f}(\vec{x})$ is the analysis of the variational equation around a particular solution $\bar{x}=\bar{x}(t)$.  The variational equation around $\bar{x}(t)$ is a linear system obtained by linearizing the vector field around $\bar{x}(t)$. If the nonlinear system admits some first integrals so does the variational equation. Thus, proving that the variational equation does not admit any first integral within a given class of functions implies that the original nonlinear system is nonintegrable. In particular one works in the analytic setting where inverting the solution $\bar{x}(t)$ one obtains a (noncompact) Riemann surface $\Gamma$ given by integrating $dt=dw/\dot{\bar{x}}(w)$ with the appropriate limits. Linearizing the system of differential equations around the straight line solution yields the {\it Normal Variational Equation } (NVE), which is the component of the linearized system which describes the variational normal to the surface $\Gamma$.

The methods described here are useful for Hamiltonian systems, luckily for us, the Virasoro constraints in string theory provide a Hamiltonian for the systems we consider. This is particularly interesting as the origin of this constraint is strictly stringy but allows a very intuitive interpretation from the dynamical system perspective. Given a Hamiltonian system, the main statement of Ziglin's theorems is to relate the existence of a first integral of motion with the monodromy matrices around the straight line solution \cite{ZiglinI, ZiglinII}. The simplest way to compute such monodromies is by changing coordinates to bring the normal variational equation into a known form (hypergeometric, Lam\'e, Bessel, Heun, etc). Basically one needs to compute the monodromies around the regular singular points, for example in the Gauss hypergeometric equation $z(1-z)\xi'' +(3/4)(1+z) \xi' +(a/8) \xi=0$, the monodromy matrices can be expressed in terms of the product of monodromy matrices obtained by taking closed paths around $z=0$ and $z=1$. In general the answer depends on the parameters of the equation, for example $a$ above.

Morales-Ruiz and Ramis proposed a major improvement on Ziglin's theory by introducing techniques of differential Galois theory \cite{MR-S,MR-R,MR-R-S}. The key observation is to change the formulation of integrability from a question of monodromy to a question of the nature of the Galois group of the NVE. Intuitively, if we go back to Kovalevskaya we are interested in understanding whether the KAM tori are resonant or not resonant or, in simpler terms, if their characteristic frequencies are rational or irrational (see the pedagogical introductions provided in \cite{Morales-Ruiz,MR-Kovalevskaya}). This statement turns out to be dealt with most efficiently in terms of the Galois group of the NVE. The key result is now stated as: If the differential Galois group of the NVE is non-virtually Abelian, that is, the identity connected component is a non-Abelian group, then the Hamiltonian system is non-integrable. The calculation of the Galois group is rather intricate, as was the calculation of the monodromies, but the key simplification comes through the application of Kovacic's algorithm \cite{Kovacic}. Kovacic's algorithm, an algorithmic implementation of Picard–Vessiot theory for second order homogeneous linear differential equations with polynomial coefficients,  gives a constructive answer to the existence of integrability by quadratures. Fortunately Kovacic's algorithm is implemented in most computer algebra software including Maple and Mathematica. It is a little tedious but straightforward to go through the steps of the algorithm manually. So, once we write down our NVE in a suitable linear form it becomes a simple task to check their solvability in quadratures. An important property of the Kovacic's algorithm is that it works if and only if the system is integrable, thus a failure of completing the algorithm equates to a proof of non-integrability. This route of declaring systems non-integrable has been successfully applied to various situations, some interesting examples include: \cite{Morales-Ramis, Mciejewski,Primitivo, lakatos}. See also \cite{ZiglinABC} for nonintegrability of generalizations of the H\'enon-Heiles system \cite{MR-Kovalevskaya}. A nice compilation of examples can be found in  \cite{Morales-Ruiz}.

\section{Wrapped strings in general $AdS_5\times X^5$}\label{sec:Models}

The methods of analytic  non-integrability can be applied to a large class of spaces in string theory. Let us start by considering a five-dimensional  Einstein space $X_5$, with $R_{ij}=\lambda g_{ij}$, where $\lambda$ is a constant.  Any such Einstein space furnishes a solution to the type IIB supergravity equations known as a  Freund-Rubin compactification \cite{Freund:1980xh}. Namely, the solution takes the form
\bea
\label{eq:Freund-Rubin}
ds^2&=&ds^2(AdS_5) + ds^2(X_5), \qquad F_5=(1+\star){\rm vol}(AdS_5),
\eea
where ${\rm vol}$ is the volume five-form and $\star$ is the Hodge dual operator. Basically $F_5$ is a the sum of the volume forms on $AdS_5$ and the Einstein space $X_5$. Of particular interest in string theory is the case when $X_5$ is Sasaki-Einstein, that is, on top of being Einstein it admits a spinor satisfying $\nabla_\mu \epsilon \sim \Gamma_\mu \epsilon$. The case of Sasaki-Einstein structure is particularly interesting from the string theory point of view as it preserves supersymmetry which is a mechanism that provides extra computational power.

We consider spaces $X_5$ that are a $U(1)$ fiber over a four-dimensional manifold, again, this is largely inspired by the Sasaki-Einstein class but clearly goes beyond that.  In the case of topologically trivial fibration we are precluded from applying our argument, those manifolds can be considered separately and probably on a case by case basis.

The configuration that we are interested in exploring is a string sitting at the center of $AdS_5$ and winding in the {\it circles} provided by the base space. More explicitly, consider the $AdS_5$ metric in global coordinates
\bea
ds^2&=& -\cosh^2\rho\,\, dt^2 + d\rho^2 + \sinh^2\rho\, d\O_3^2.
\eea
Then, our solutions is localized at $\rho=0$. As noted before, in the search for solutions of the form (\ref{eq:Freund-Rubin}) a prominent place is taken by deformations of $S^5$ that preserve some amount of supersymmetry, they are given by Sasaki-Einstein spaces. The general structure of Sasaki-Eintein metrics is
\be
ds^2_{X^5_{S-E}}= (d\psi+ \frac{i}{2}(K_{,i} dz^i - K_{,\bar{i}}d\bar{z}^i))^2+ K_{,i\bar{j}} dz^id\bar{z}^j,
\ee
where $K$ is a K\"ahler potential on the complex base with coordinates $z_i$ with $i=1,2$. This is the general structure that will serve as our guiding principle but we will not be limited to it. Roughly our Ansatz for the classical string configuration is
\bea
z_i=r_i(\tau) e^{i\alpha_i \sigma}
\eea
where $\tau$ and $\sigma$ are the worldsheet coordinates of the string. Note crucially we have introduced winding of the strings characterized by the constants $\alpha_i$. The goal is to solved for the functions $r_i(\tau)$.

A summary from the previous section instructs us to:

\begin{itemize}

\item Select a particular solution, that is, define the {\it straight line solution}.

\item Write the  normal variational equation (NVE).

\item Check if the identity component of the differential Galois group of the NVE is Abelian, that is, apply the Kovacic's algorithm to determine if the NVE is integrable by quadrature.

\end{itemize}

Given this Ansatz above we can now summarize the general results. Namely, in this section we prove that the corresponding effective Hamiltonian systems have two degrees of freedom and admit an invariant plane $\Gamma=\{r_2=\dot{r}_2=0\}$ whose normal variational equation around integral curves in $\Gamma$ we study explicitly.

\subsection{$T^{p,q}$}

These 5-manifolds are not necessarily Sasaki-Einstein, however, some of them are Einstein which allow for a consistent string backgrounds of the form described in equation (\ref{eq:Freund-Rubin}). More importantly, some of these spaces provide exact conformal sigma models description \cite{PandoZayas:2000he} and are thus exact string backgrounds in all orders in $\alpha'$. In this section we provide a unified treatment of this class for generic values of $p$ and $q$. The metric has the form
\be
ds^2=a^2 (d\psi +p\cos\theta_1d\phi_1+q\cos\theta_2d \phi_2)^2 +b^2 (d\theta_1^2+\sin^2\theta_1 d\phi_1^2) +
c^2 (d\theta_2^2+\sin^2\theta_2 d\phi_2^2).
\ee
The classical string configuration we are interested is
\bea
\theta_1&=&\theta_1(\tau), \qquad \theta_2=\theta_2(\tau), \qquad \psi=\psi(\tau), \qquad t(\tau), \nonumber \\
\phi_1&=&\alpha_1 \sigma, \qquad \phi_2=\alpha_2 \sigma,
\eea
where $\alpha_i$ are constants quantifying how the string wounds along the $\phi_i$ directions. Recall that $t$ is from AdS$_5$. The Polyakov Lagrangian is
\bea
{\cal L}&=&-\frac{1}{2\pi\alpha'}\bigg[\dot{t}^2 -b^2 \dot{\theta}_1^2- -c^2 \dot{\theta}_2^2-a^2 \dot{\psi}^2 \nonumber \\
&+& \alpha_1^2 (b^2-a^2p^2)\sin^2\theta_1 +  \alpha_2^2 (c^2-a^2q^2)\sin^2\theta_2 +2\alpha_1\alpha_2 p\, q\, a^2 \cos\theta_1\cos\theta_2\bigg].
\eea

There are several conserved quantities, the corresponding nontrivial equations are
\bea
\label{eq:TpqSystem}
\ddot{\theta}_1 &+& \frac{\alpha_1}{b^2}\sin\theta_1\bigg[\alpha_1 (b^2-a^2p^2)\cos\theta_1 - a^2 \alpha_2 p q \cos\theta_2 \bigg]=0, \nonumber \\
\ddot{\theta}_2 &+& \frac{\alpha_2}{c^2}\sin\theta_2\bigg[\alpha_2 (c^2-a^2q^2)\cos\theta_2 - a^2 \alpha_1 p q \cos\theta_2 \bigg]=0.
\eea
There is immediately some insight into the role of the fibration structure.  Note that the topological winding in the space which is roughly described by  $p$ and $q$ intertwines with the wrapping of the strings $\alpha_1$ and $\alpha_2$. The effective number that appears in the interaction part of the equations is $\alpha_1\, p$ and $\alpha_2 q$. For example, from the point of view of the interactions terms,  taking $p=0$ or $q=0$ is equivalent to taking one of the $\alpha_i=0$ which  leads to an integrable system of two non-interacting gravitational pendula.

Following the structure of the discussion of section \ref{sec:Analytic}, we take the straight line solution to be
\be
\theta_2=\dot{\theta}_2=0.
\ee
The equation for $\theta_1$ becomes
\be
\ddot{\theta}_1 +\frac{\alpha_1}{b^2}\bigg[\alpha_1 (b^2 -a^2p^2)\cos\theta_1 -a^2\alpha_2 pq\bigg]\sin\theta_1=0.
\ee
Let us denote the solution to this equation $\bar{\theta}_1$, it can be given explicitly but we will not need the precise form. This solution also defines the Riemann surface $\Gamma$ introduced in section \ref{sec:Analytic}. The NVE is obtained by considering small fluctuations in $\theta_2$ around the above solutions and takes the form:
\be
\ddot{\eta}+\frac{\alpha_2}{c^2}\bigg[\alpha_2 (c^2-a^2 q^2) -\alpha_1 p\, q \cos\bar{\theta}_1\bigg]\eta=0.
\ee
Our goal is to study the NVE. To make the equation amenable to the Kovacic's algorithm we introduce the following substitution
\be
\cos(\bar{\theta}_1)=z.
\ee
In this variable the NVE takes a form similar to Lam\'e equation (see section 2.8.4 of \cite{Morales-Ruiz}),
\begin{align}
\label{eq:TpqNVE}
f(z) \eta''(z)+\frac{1}{2} f'(z) \eta'(z)+\frac{\alpha_2}{c^2}\bigg[\alpha_2 (c^2-a^2 q^2) -\alpha_1 p\, q z \bigg] \eta(z) =0
\end{align}
where, prime now denotes differentiation with respect to $z$.
\begin{align}
f(z)=\dot{\bar{\theta}}_1^2 \sin^2(\bar{\theta}_1)=\left(6 E^2 -\frac{1}{3} (4 \alpha_1\alpha_2 z+\alpha_2^2 (1-z^2))\right)(1-z^2)
\end{align}
Equation (\ref{eq:TpqNVE}) is a second order homogeneous linear differential equation with polynomial coefficients and it is, therefore, ready for the application of Kovacic's algorithm. For generic values of the parameters above the Kovacic's algorithm does no produce a solution meaning the system defined in equations (\ref{eq:TpqSystem}) is not integrable.

\subsection{NVE for $T^{1,1}$}
It is worth taking a pause to discussed the case of $T^{1,1}$ explicitly. The NVE equation takes a simpler form:
\begin{align}
\ddot \eta+\frac{1}{3}(\alpha_1^2-2\alpha_1\alpha_2 \cos(\theta_1(t))\eta =0 ,
\end{align}
where $\eta$, as above, is the variation in $\theta_2$. Substituting $\cos(y)=z$ this equation takes a form similar to Lam\'e equation
\begin{align}
f(z) \eta''(z)+\frac{1}{2} f'(z) \eta'(z)+\frac{1}{3}(\alpha_1^2-2\alpha_1\alpha_2 z )\eta(z) =0
\end{align}
Similarly we can obtain an expression for the function $f(z)$ as
\begin{align}
f(z)=\dot y^2 \sin^2(y)=\left(6 E^2 -\frac{1}{3} (4 \alpha_1\alpha_2 z+\alpha_2^2 (1-z^2))\right)(1-z^2).
\end{align}
Consequently, this system is also non-integrable.

The case of $T^{1,1}$ is particularly interesting because in the case the supergravity solution is supersymmetric and a lot of attention has been paid to extending configurations of $AdS_5\times S^5$ to the case of $AdS_5\times T^{1,1}$  \cite{PandoZayas:2002rx,Itzhaki:2002kh,Gomis:2002km,Kim:2003vn,Wang:2005baa,Benvenuti:2005cz,Benvenuti:2007qt}.

\subsection{$Y^{p,q}$}

These spaces have played a central role in developments of the AdS/CFT correspondence as they provided and infinite class of dualities. These spaces are Sasaki-Einstein but they are not coset spaces as was the case for the $Y^{p,q}$ discussed above.  Following the general discussion of Sasaki-Einstein spaces above, we write the metric on these spaces as

\be
\frac{1}{R^2}ds^2=\frac{1}{9}(d\psi -(1-cy) \cos\theta d\phi +y d\beta)^2 +\frac{1-cy}{6} (d\theta^2+\sin^2\theta d\phi^2) +
\frac{p(y)}{6}(d\beta +c\, \cos\theta d\phi)^2.
\ee
\be
p(y)=\frac{a-3y^2 +2c\, y^3}{3(1-c\, y)}.
\ee
The classical string configuration is described by the Ansatz
\bea
\theta&=&\theta (\tau), \qquad y=y(\tau), \nonumber \\
\phi&=&\alpha_1 \sigma, \qquad \beta=\alpha_2 \sigma,
\eea
The Polyakov Lagrangian is simply:
\bea
{\cal L}&=&-\frac{1}{2\pi\alpha'}\bigg[\dot{t}^2 -\frac{1-cy}{6} \dot{\theta}^2-\frac{1}{6p(y)}\dot{y}^2 -\frac{1}{9} \dot{\psi}^2 \nonumber \\
&+& \frac{1-cy}{6}\alpha_1^2 \sin^2\theta +  \frac{p(y)}{6} (\alpha_2 +c\, \alpha_1 \cos\theta)^2
+\frac{1}{9}(\alpha_2\, y -\alpha_1(1-cy)\cos\theta)^2\bigg]
\eea
As in previous cases the equations of motion for $t$ and $\psi$ are integrated immediately leaving only two nontrivial equations for $\theta$ and $y$
\bea
\ddot{\theta}&-&\frac{c}{1-c\, y} \dot{y}\dot{\theta}+\alpha_1\left(\alpha_1\cos\theta -\frac{c\, p(y)}{1-c\, y}(\alpha_2+c\, \alpha_1 \cos\theta)-\frac{2}{3}(\alpha_2 \, y -\alpha_1(1-c\, y)\cos\theta)\right)\sin\theta=0. \nonumber \\
\ddot{y}&-&\frac{p'}{p}\dot{y}^2 +\frac{p\, p'}{2}(\alpha_2 +c\, \alpha_1 \cos\theta)^2 -\frac{c\, p}{2}\alpha_1^2\sin^2\theta
+\frac{2}{3}p(\alpha_2 +c\, \alpha_1 \cos\theta)(\alpha_2 \, y -\alpha_1(1-c\, y)\cos\theta)=0. \nonumber
\eea

\subsubsection{$\theta$ straight line}

The straight line solution ca be taken to be $\theta=\dot{\theta}=0$. Then the equation for $y$ is simplified to
\be
\ddot{y}-\frac{p'}{p}\dot{y}^2 +\frac{p\, p'}{2}(\alpha_2 +c\, \alpha_1 )^2 +\frac{2}{3}p(\alpha_2 +c\, \alpha_1 )(y(\alpha_2 +c\, \alpha_1)-\alpha_1)=0.
\ee
The Normal Variational Equation takes the form
\be
\ddot{\eta}-\frac{c\, \dot{y}_s }{1-c y_s}\dot{\eta}+\alpha_1 \left(\alpha_1 -\frac{c\, p(y_s)}{1-c\, y_s}(\alpha_2 +c\, \alpha_1)
-\frac{2}{3}( (\alpha_2 + c\, \alpha_1)y_s-\alpha_1)\,\right)\eta=0.
\ee
To be able to write it in a form conducive to the application of Kovacic's algorithm we subtitute $y_s(t)=y$ and the NVE takes the form
\bea
\left(\ddot{y}(t)(1-cy)-c \dot{y}^2(t)\right) \frac{d \eta}{d y}+(1-c y)\dot{y}^2(t) \frac{d^2 \eta}{d y^2}+q(y) n(y)=0 \\
\nonumber \dot{y}^2(t)=6(E+p(y) V(y,0))=6 p(y) \left(\frac{p(y)}{6}(\alpha_2+c \alpha_1)^2+\frac{1}{9}(\alpha_2 y-\alpha_1(1-c y))\right) \\
\nonumber \ddot{y}(t)= 3 \frac{d}{d y}( p(y) V(y,0) )\\
\nonumber q(y)= \alpha_1 \left( 1-cy \right)  \left( 5/3\,{\it \alpha_1}-{\frac {c \left(
a-3\,{y}^{2}+2\,c{y}^{3} \right)  \left( {\alpha_2}+c{\it \alpha_1} \right) }{
 \left( 3-3\,cy \right)  \left( 1-cy \right) }}-2/3\, \left( {\alpha_2}+
c{ \alpha_1} \right) y \right)
\eea
With this identifications we have rewritten the NVE as a homogeneous second order linear differential equation. The Kovacic's algorithm again fails to yield a solution pointing to the fact that the system is generically non-integrable.

\subsection{The exceptional case: $S^5$}
In this section we provide an integrable example where the Kovacic's algorithm should succeed. To expose the Sasaki-Einstein structure of $S^5$, it is convenient to write the metric as a $U(1)$ fiber over $\mathbb{P}^2$. The round metrics on  $S^{5}$ may be elegantly expressed in terms of the left-invariant one-forms of $SU(2)$. For $SU(2)$, the left-invariant one-forms  can be written  as,

\bea
\sigma_1 &=&\frac12(\cos(d\psi )d\theta  + \sin(\psi) \sin(\theta) d\phi), \nonumber \\
\sigma_2 & =&\frac12 ( \sin(\psi) d\theta -  \cos(\psi) \sin(\theta)d \phi), \nonumber \\
\sigma_3 &=&\frac12(d\psi  + \cos(\theta)d\phi).
\eea
In terms of these 1-forms, the metrics on $\mathbb{P}^2$ and $S^5$ may be written,
\bea
ds^2_{\mathbb{P}^2} &=& d\mu^2 + \sin^2(\mu)\left(\sigma^2_1 + \sigma^2_2 + \cos^2(\mu)\sigma^2_3\right), \nonumber \\
ds^2_{S^5} &=& ds^2_{\mathbb{P}^2} +(d\chi + \sin^2(\mu)\sigma_3)^2
\eea
where $\chi$  is the local coordinate on the Hopf fibre and $A = \sin^2(\mu)\sigma_3 = \sin^2(\mu)(d\psi + \cos(\theta)d\phi)/2$
is the 1-form potential for the K\"ahler form on $\mathbb{P}^2$ \cite{Adams:2008wt}.

The classical string configuration is
\bea
\theta&=&\theta (\tau), \qquad \mu=\mu(\tau),  \chi=\chi(\tau), \nonumber \\
\phi&=&\alpha_1 \sigma, \qquad \psi=\alpha_2 \sigma,
\eea

The Lagrangian is
\bea
{\cal L}&=&-\frac{1}{2\pi\alpha'}\bigg[\dot{t}^2 -\dot{\mu}^2 -\frac{1}{4}\sin^2\mu \dot{\theta}^2 -\dot{\chi}^2 \nonumber \\
&+& \frac{1}{4}\sin^2\mu \left(\alpha_1^2 \sin^2\theta+(\alpha_2+\alpha_1 \cos\theta)^2\right)\bigg].
\eea
The nontrivial equations of motion are
\bea
\ddot{\mu}&+&\frac{1}{8}\sin(2\mu)\bigg[\dot{\theta}^2 -2\alpha_1\alpha_2 \cos\theta -\alpha_1^2 -\alpha_2^2\bigg]=0, \nonumber \\
\ddot{\theta}&+&2\dot{\mu}\dot{\theta}\cot(\mu) +\alpha_1\alpha_2 \sin\theta=0.
\eea
Inspection of the above system shows that we have various natural choices. We discussed the two natural choices of straight line solutions in what follows.
\subsubsection{$\theta$ straight line}
Let us assume $\theta=\dot{\theta}=0$, then the equation for $\mu$ becomes
\be
\ddot{\mu}-\frac{1}{8}(\alpha_1+\alpha_2)^2 \sin (2\mu)=0.
\ee
We call the solution of this equation $\mu_s$. The NVE is
\be
\ddot{\eta}+2\cot(\mu_s)\dot{\mu}_s \dot{\eta}+\alpha_1\alpha_2 \eta=0.
\ee
With $\sin(\mu)=z$ the NVE may be written as,
\begin{align}
p(z){\frac {d^{2}}{d{z}^{2}}}\eta & \left( z \right)+q(z){\frac {d}{d{z}}}\eta \left( z \right)+\alpha_1\alpha_2 z^2 \eta(z)=0 \\
p(z)& ={z}^{2} \left( 2 E+1/8 \left( { \alpha_1}+{ \alpha_2} \right) ^{2}
 \left( 1-2{z}^{2} \right)  \right)  \left( 1-{z}^{2} \right)   \\
\nonumber q(z)&=-1/8z \big( -32 E+48 E z^{2}-2 \alpha_1^2+9
{ \alpha_1}^{2}{z}^{2}-8{\alpha_1}^2 z^{4}-4{\alpha_1}{\alpha_2} \\
\nonumber & + 18{\alpha_1}{\alpha_2}{z}^{2}-16{\alpha_1}{ \alpha_2}{z}^{4}-2{
 \alpha_2}^{2}+ 9{\alpha_2}^{2}{z}^{2}-8{\alpha_2}^{2}{z}^{4}
 \big)
\end{align}
This equation is now on the form conducive to Kovacic's algorithm which succeeds and gives a solution. Since the above approach obscure the nature of integrability of $AdS_5\times S^5$ we consider another example which leaves no doubt about the integrability.

\subsubsection{$\mu$ straight line}

Let us assume the straight line is now given by $\mu=\pi/2, \dot{\mu}=0$. The the equation for $\theta$ becomes
\be
\ddot{\theta}+\alpha_1\alpha_2 \sin\theta=0.
\ee
Let us call the solution to this equation $\theta_s$. Then the NVE is
\be
\ddot{\eta}+\frac14\left(\dot{\theta}_s^2-2\alpha_1 \alpha_2 \cos(\theta_s)-\alpha_1^2 -\alpha_2^2\right)\eta=0.
\ee

Note that the equation of motion for $\theta_s$ implies
\bea
&&\ddot{\theta}+\alpha_1\alpha_2 \sin\theta=0 \nonumber \\
&\rightarrow &\frac{d}{d\tau}\left(\dot{\theta}_s^2-2\alpha_1\alpha_2\cos\theta_s\right)=0, \nonumber \\
&\rightarrow & \dot{\theta}_s^2-2\alpha_1\alpha_2\cos\theta_s=C_0
\eea

Thus the NVE equation can be written as a simple harmonic equation
\be
\ddot{\eta}+\frac14\left(C_0-\alpha_1^2 -\alpha_2^2\right)\eta=0.
\ee
We do not require Kovacic's algorithm to tell us that there is an analytic solution for this equation. The power of differential Galois theory also guarantees that the result is really independent of the straight line solution (Riemann surface) that one chooses.

We conclude this subsection with the jovial comment that we now know a very precise sense in which {\it String theory in $AdS_5\times S^5$ is like a harmonic oscillator.}

\section{ Conclusions }\label{sec:conclusions}
In this paper we have shown that certain classical string configurations corresponding to a string winding along two of the angles of a general class of five-dimensional Einstein manifolds $X_5$, realized as a nontrivial $S^1$ fibration over a 4-d base, is non-integrable.

In all the previous examples in the mathematical literature homogeneity of the potential played a crucial role in the proof  \cite{Morales-Ramis,lakatos,Mciejewski}. Another mathematical curiosity comes from the fact that traditionally due to the works of Hadamard and later of Anosov, chaos has been associated with the motion of particles in negatively curved spaces through the Jacobi equation. The class of five-dimensional Einstein spaces used here have  positive curvature. The main mechanism for non-integrability is provided by winding of the strings which is a property unique to strings and therefore not well understood. More precisely, we found an interesting interplay topology $c_1=\int dA$ and dynamics as the Chern class determines the possibility of an interaction term in the dynamical system. As pointed out in the main terms in various cases the interaction and therefore the non-integrability appears as the product of the Chern number and the winding number of the string.

The direct connection between analytic non-integrability and chaotic behavior is still open. Of course, the systems that have been proven to be non-integrable using differential Galois theory were suspected to be chaotic already. More prominently is the prototypical case of Henon-Heiles. This question has been discussed in the literature and we refer the reader to \cite{Morales-Ruiz} for further details. For the safe of disclosure we notice that we have not directly proved that the systems discussed in section \ref{sec:Analytic} are chaotic. However, together with our previous publication \cite{Basu:2011di}, we believe the case for outright chaotic behavior is overwhelmingly strong.

\section*{Acknowledgments}

PB thanks Diptarka Das, Sumit Das, Archisman Ghosh and Al Shapere. This work is  partially supported by Department of Energy under grant DE-FG02-95ER40899 to the University of Michigan. PB is partially supported by a National Science Foundation grant NSF-PHY-0855614.

\end{document}